\documentclass[aps,pre,twocolumn,groupedaddress,showpacs,floatfix,superscriptaddress]{revtex4-1}
\usepackage{epsfig}
\usepackage{amsmath,amssymb}
\usepackage{graphicx}
\usepackage[dvipsnames,usenames]{color}
\usepackage[normalem]{ulem}
\usepackage{soul}  
\usepackage[colorlinks=true, citecolor = blue]{hyperref} 
\emergencystretch=\maxdimen
\begin{document}
%%%%%%%%%%%%%%%%%%%%%%%%%%%%%%%%%%%%%%%%%%
\title{Giant spin current rectification due to the interplay of negative differential conductance and a non-uniform magnetic field}

\author{Kang Hao Lee}
\affiliation{Science and Math Cluster, Singapore University of Technology and Design, 8 Somapah Road, 487372 Singapore}

\author{Vinitha Balachandran}
\email{vinitha\_balachandran@sutd.edu.sg}
\affiliation{Science and Math Cluster, Singapore University of Technology and Design, 8 Somapah Road, 487372 Singapore}

\author{Ryan Tan}
\affiliation{Engineering Product Development Pillar, Singapore University of Technology and Design, 8 Somapah Road, 487372 Singapore}

\author{Chu Guo}
\affiliation{Key Laboratory of Low-Dimensional Quantum Structures and Quantum Control of Ministry of Education, Department of Physics and Synergetic Innovation Center for Quantum Effects and Applications, Hunan Normal University, Changsha 410081, China}

\author{Dario Poletti}
\email{dario\_poletti@sutd.edu.sg}
\affiliation{Science and Math Cluster, Singapore University of Technology and Design, 8 Somapah Road, 487372 Singapore}
\affiliation{Engineering Product Development Pillar, Singapore University of Technology and Design, 8 Somapah Road, 487372 Singapore}

\date{\today}
\begin{abstract}
In XXZ chains with large enough interactions, spin transport can be significantly suppressed when the bias of the dissipative driving becomes large enough. This phenomenon of negative differential conductance is caused by the formation of two oppositely polarized ferromagnetic domains at the edges of the chain. Here we show that this many-body effect, combined with a non-uniform magnetic field, can allow a high degree of control of the spin current.
In particular, by studying all the possible shapes of local magnetic fields potentials, we found that a configuration in which the magnetic field points up for half of the chain and down for the other half, can result in giant spin-current rectification, for example up to $10^8$ for a system with only $8$ spins. Our results show clear indications that the rectification can increase with the system size.
\end{abstract}
\maketitle
%%%%%%%%%%%%%%%%%%%%%%%%%%%%%%%%%%%%%%%%%%

\section{\label{sec:level1}Introduction}
Quantum spin systems exhibit rich transport properties. For instance, tuning the interactions in the system, spin transport can change from ballistic to diffusive \cite{prosen_open_2011, znidaric_spin_2011, prosen_exact_2011, BertiniZnidaric2020}. One effect that is particularly relevant for our work is the emergence of negative differential conductance (NDC), that is the phenomenon whereby the spin current decreases as the bias imposed by the spin baths increases \cite{benenti_charge_2009, benenti_negative_2009}. Such an apparently counterintuitive phenomenon is due to the fact that the interplay between the dissipative driving and the interactions in the system result in the formation of ferromagnetic domains at the edges of the chain, which significantly suppress the spin current. The effect can be so strong that the spin chain becomes an insulator.

Here we study a boundary driven XXZ spin chain in the NDC regime in the presence of a non-uniform external magnetic field. In order to obtain more generic conclusions we consider the magnetic field that locally can only take two possible values $\pm h$.
A detailed analysis of the effect of different shapes of the magnetic field potential show that two configurations, such that the magnetic field is in one direction in half of the chain and in the other direction for the other half, strongly enhance or even more strongly suppress the ferromagnetic domains. This results, respectively, in the smallest or largest spin currents between all the possible shapes of the magnetic field potential. Since these two configurations are mirror-symmetric, this implies that if the field points for half the chain in one direction, and for the other half in the opposite direction, then one can obtain a giant rectification effect which, we show, can be of the order of $10^8$ already for small spin chains. 
The currents and rectification also show a resonant behavior which we correlate to the presence of avoided crossings in the energy spectrum of the bulk Hamiltonian. An analysis of the delocalization of the eigenstates of the Hamiltonian indicates that this giant rectification is present also in the thermodynamic limit.

This work adds to the recent results on rectification in spin chains without local magnetic fields \cite{balachandran_perfect_2018, BalachandranPoletti2019a}, with disorder \cite{BalachandranPoletti2019} or with external fields \cite{ArracheaAligia2009, SchuabLandi2016, landi_open_2015, Pereira2019, Pereira2019b, Oliveira2020}.
The manuscript is organized as follows: in Sec.\ref{sec:model} we describe our model, and in Sec.\ref{sec:results} we discuss our results. Last, in Sec.\ref{sec:conclusions} we draw our conclusions.

%%%%%%%%%%%%%%%%%%%%%%%%%%%%%%%%%%%%%%%%%%
\section{Model}\label{sec:model}
We consider an XXZ spin chain of length $L$ with the following Hamiltonian
\begin{equation}
\hat{\mathcal{H}}=\sum_{i=1}^{L-1} 2J(\hat{\sigma}^+_i\hat{\sigma}^-_{i+1} + \hat{\sigma}^-_i\hat{\sigma}^+_{i+1}) + J_{zz}\hat{\sigma}^z_i\hat{\sigma}^z_{i+1} + \sum_{i=1}^{L}h_i\hat{\sigma}^z_i,
\label{ham}
\end{equation}
where $\hat{\sigma}^{\pm}_i$ are the raising and lowering operators acting on site $i$ and $\hat{\sigma}^z_i$ is a Pauli spin matrix. $J$ and $J_{zz}$ denote the tunneling strength and magnitude of the nearest neighbor interaction respectively. We use $h_i$ for the local magnetic field. On each site, the local magnetic field $h_i$ can take only the two discrete values $\pm h$. Therefore, there are $2^L$ possible shapes of magnetic field potential.

The chain is coupled to two spin baths at the edges and we model the evolution via a Gorini-Kossakowski-Sudarshan-Lindblad (GKSL) master equation \cite{gorini_completely_1976,lindblad_generators_1976} for the system density matrix as \cite{MendozaArenasClark2013,mendoza-arenas_heat_2013,znidaric_exact_2010,znidaric_transport_2013,prosen_exact_2011,karevski_exact_2013,landi_open_2015,prosenmatrix2009,znidaric_dephasing-induced_2010}
\begin{equation}
\frac{\partial\hat{\rho}}{\partial t}=-\frac{\rm{i}}{\hbar}[\hat{\mathcal{H}},\hat{\rho}]+\sum^4_{n=1}\hat{\Gamma}_n\hat{\rho} \;\hat{\Gamma}_n^\dagger-\frac{1}{2}\sum^4_{n=1}\{\hat{\Gamma}^\dagger_n \hat{\Gamma}_n,\hat{\rho}\},
\label{lindblad}
\end{equation}
where the $\hat{\Gamma}_n$ are the jump operators given by
\begin{align}
    \hat{\Gamma}_1 &= \sqrt{\gamma\mu_L}\hat{\sigma}_1^+, & \hat{\Gamma}_2 &=\sqrt{\gamma(1-\mu_L)}\hat{\sigma}_1^-,\\
    \hat{\Gamma}_3 &= \sqrt{\gamma\mu_R}\hat{\sigma}_N^+, & \hat{\Gamma}_4 &=\sqrt{\gamma(1-\mu_R)}\hat{\sigma}_N^-. \label{eq: set of dissipators}
\end{align}

Here $\gamma$ describes the system-reservoir coupling strength and $\mu_L$ ($\mu_R$) is the left (right) dissipation bias. We choose a symmetric driving at the boundaries, i.e. $\mu_{L,R} = (1\mp\mu)/2$. Thus, $\mu \equiv \mu_R-\mu_L \in[-1,1]$ is the dissipative boundary driving bias due to the reservoirs. In the limiting case with $\mu=1$ so that $\mu_L = 0$ and $\mu_R=1$, the left reservoir tries to impose spin up to spin down conversions, while the right reservoir would do the opposite, only converting spins down to spins up. For the rest of the paper, we will be using $\mu=1$, which is the strongest possible bias in the study of our systems.

For $\mu\neq0$, the system relaxes to a current carrying non-equilibrium steady state (NESS) $\hat{\rho}_{ss}$ at long times. The spin current $\mathcal{J}$ can be obtained from the continuity equation for local magnetisation $ \hat{\sigma}_i^z$ as $\mathcal{J} = \textrm{Tr}(\hat{j}_i \hat{\rho}_{ss})$, where $\hat{j}_{i}  = 4\mathrm{i}J( \hat{\sigma}^-_i \hat{\sigma}^+_{i+1} - \hat{\sigma}^+_i \hat{\sigma}^-_{i+1})/\hbar$. In the steady state, the current is independent of the chosen site $i$. For all systems considered in this study,  $\hat{\rho}_{ss}$, is computed by setting the time derivative to zero in Eq.(\ref{lindblad}) and using exact diagonalization with a number conserving numerical approach described in \cite{guo_dissipatively_2017} which allows to study open spin systems up to $14$ spins. We stress however, that simulating $8$ spins for an open system with exact diagonalization corresponds to simulating $16$ spins for a Hamiltonian system. In the following we work in units for which $J$ and $\hbar$ are $1$.
%%%%%%%%%%%%%%%%%%%%%%%%%%%%%%%%%%%%%%%%%%
\section{Results}\label{sec:results}

Interactions in the XXZ chain can significantly alter the spin transport in a boundary driven chain. For instance in the absence of any field and for $\mu=1$, the spin current is ballistic for $| J_{zz}/J|<1$, super diffusive for $|J_{zz}/J|=1$ and insulating for $|J_{zz}/J|>1$. The insulating behavior at strong driving results in the interesting phenomenon of negative differential conductance in strongly interacting XXZ chains \cite{benenti_charge_2009, benenti_negative_2009}. The insulating behavior is attributed to the formation of two oppositely polarized ferromagnetic domains in the chain, each half of the chain acquiring the polarisation of the reservoir to which it is connected. The two domains inhibit the spin flips resulting in the reduction of current in the chain. The main focus of this paper is to explore the potential advantages of these ferromagnetic domains in device applications. To this end, we apply a local magnetic field in all the possible shapes of magnetic field potential configurations as presented in Eq. (\ref{ham}) to the XXZ chain and study the spin transport.
\begin{figure}[ht]
    \includegraphics[width=\columnwidth]{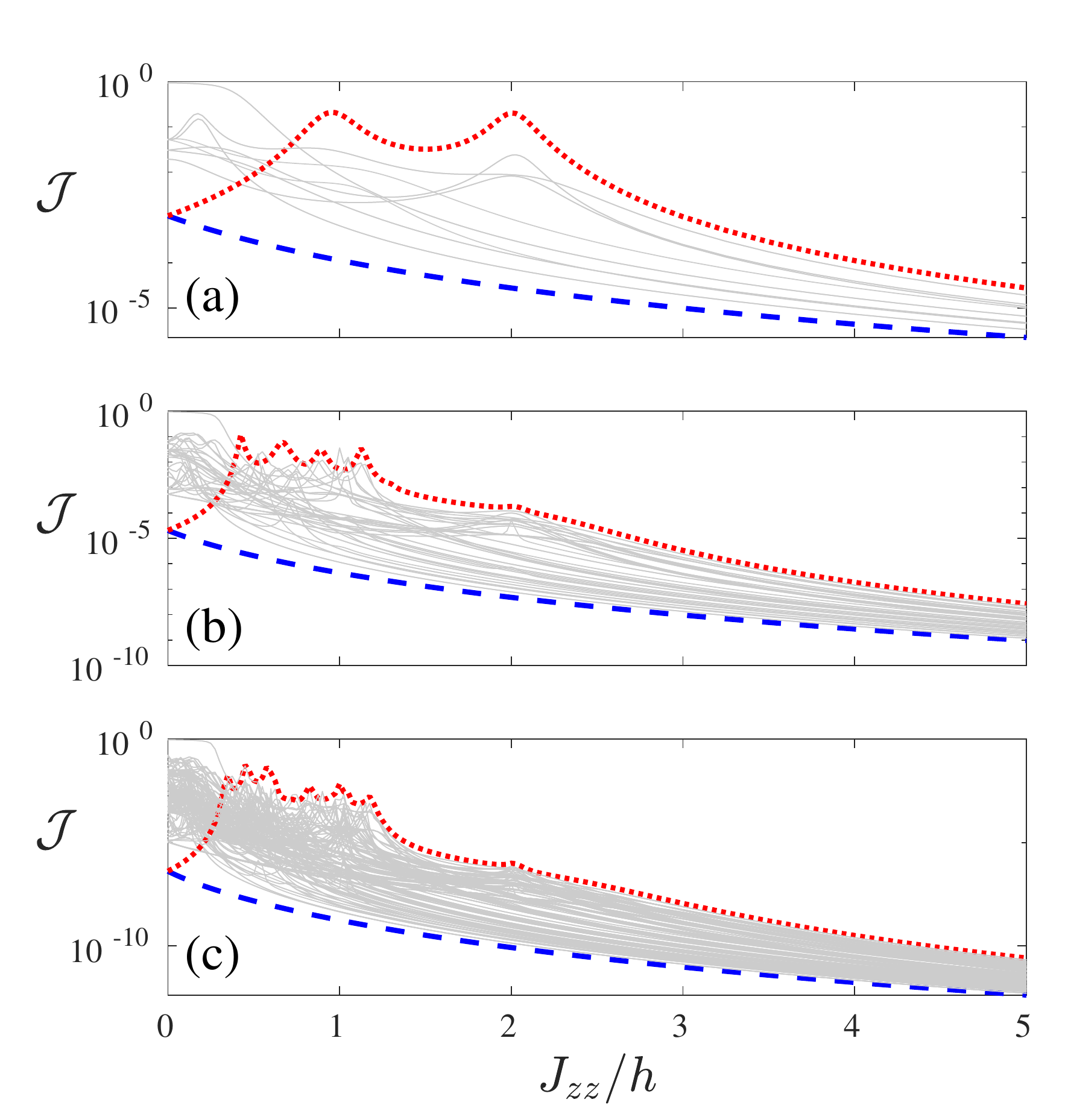}
\caption{Spin current $\mathcal{J}$ as a function of the ratio of interaction and local field strength $J_{zz}/h$ for system sizes $L=4$ (a), $L=6$ (b) and $L=8$ (c). Different lines corresponds to each of the $2^L$ magnetic field configurations.
We highlight two magnetic field configurations: with the red dotted line we show the current for a field which is $h$ for the first half of the chain, and $-h$ for the second half of the chain, which we refer to as $(+,\dots, +,-,\dots, -)$, and with the blue dashed line the realization in which the field is $-h$ in the first half of the chain and $h$ in the second half $(-,\dots, -,+,\dots, +)$.
Common parameters are $h=4$, $\gamma=1$ and $\mu=1$.}
\label{fig:currentvsdelta}
\end{figure}

We start by considering, in Fig.\ref{fig:currentvsdelta}, the spin current $\mathcal{J}$ versus interaction $J_{zz}$ for all the $2^L$ configurations of the magnetic field. In the following we use the following notation to indicate the magnetic fields direction: we write a $+$ for a site with magnetic field $+h$ and $-$ for a site with field $-h$. For instance, $(+,-,-,+)$ corresponds to the magnetic field configuration $(+h,-h,-h,+h)$. 
For the case in which the magnetic field is $h$ in the first half of the chain, and $-h$ in the second half of the chain, we refer to it as $(+,\dots +,-,\dots -)$, the magnetic field  which is $-h$ in the first half of the chain and $+h$ in the second half, we refer to it as $(-,\dots -,+,\dots +)$. Note that the configuration $(+,\dots +,-,\dots -)$  is highlighted by the red dotted line while its reflection symmetric configuration $(-,\dots -,+,\dots +)$ is depicted by the blue dashed line while all the other configurations by the grey lines. For $J_{zz}$ large enough we observe that the configurations corresponding to the blue and the red line are either the ones with the largest or the lowest currents. This is observed very clearly for system sizes $L =4$ to $8$.

\begin{figure}[ht]
    \includegraphics[width=\columnwidth]{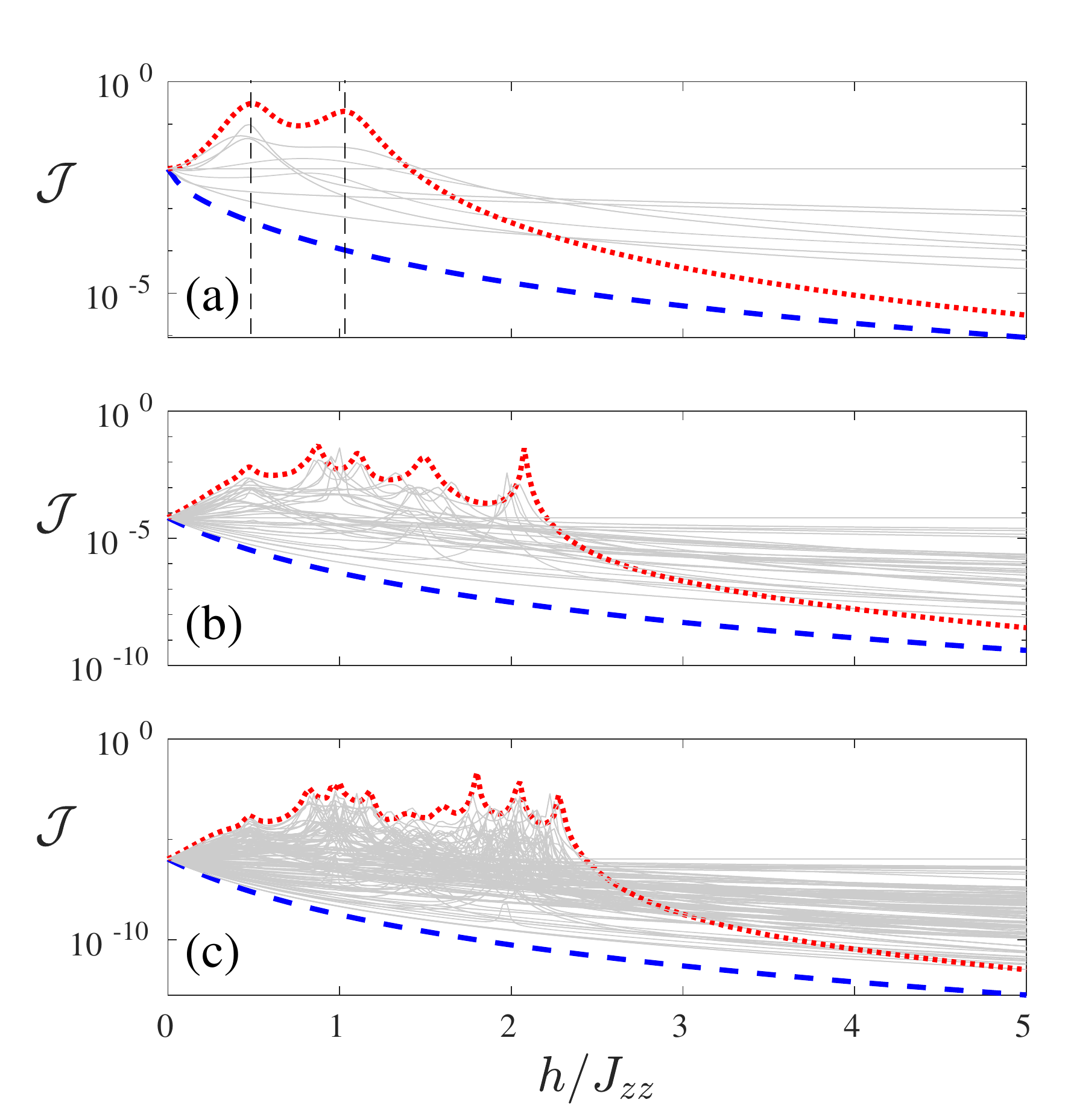}
\caption{Spin current $\mathcal{J}$ as a function of the ratio of local field strength and interaction $h/J_{zz}$  for system sizes $L=4$ (a), $L=6$ (b) and $L=8$ (c). Different lines corresponds to each of the $2^L$ configurations of local fields.
We highlight two magnetic field configurations: with the red dotted line we show the current for a field which is $h$ for the first half of the chain, and $-h$ for the second half of the chain, which we refer to as $(+,\dots, +,-,\dots, -)$, and with the blue dashed line the realization in which the field is $-h$ in the first half of the chain and $h$ in the second half $(-,\dots, -,+,\dots, +)$. Peaks of red dotted line in panel (a) are signalled by black dashed lines that correspond to the black dashed lines in Fig.\ref{fig:Fig5}.
Common parameters are $J_{zz}=4$, $\gamma=1$ and $\mu=1$. }
\label{fig:currentvsh}
\end{figure}

In Fig.\ref{fig:currentvsdelta} we consider a large local field $h=4$. It is however insightful to fix the interaction to be large, e.g. $J_{zz}=4$ and study the current as we vary $h$. This is depicted in Fig.\ref{fig:currentvsh}. The configuration $(-,-,+,+)$ corresponds (blue dashed line) to the lowest current, while the configuration $(+,+,-,-)$ corresponds, for smaller $h$, to the largest currents. It also presents some resonant-like structures, and its current decreases for larger values of $h$. Given this seemingly antithetic effect of the $(-,-,+,+)$ and $(+,+,-,-)$ configurations, which are reflection symmetric of each other, in the following we study the effectiveness of all the different magnetic field configurations to have a large spin current rectification effect.

\begin{figure}[ht]
    \includegraphics[width=\columnwidth]{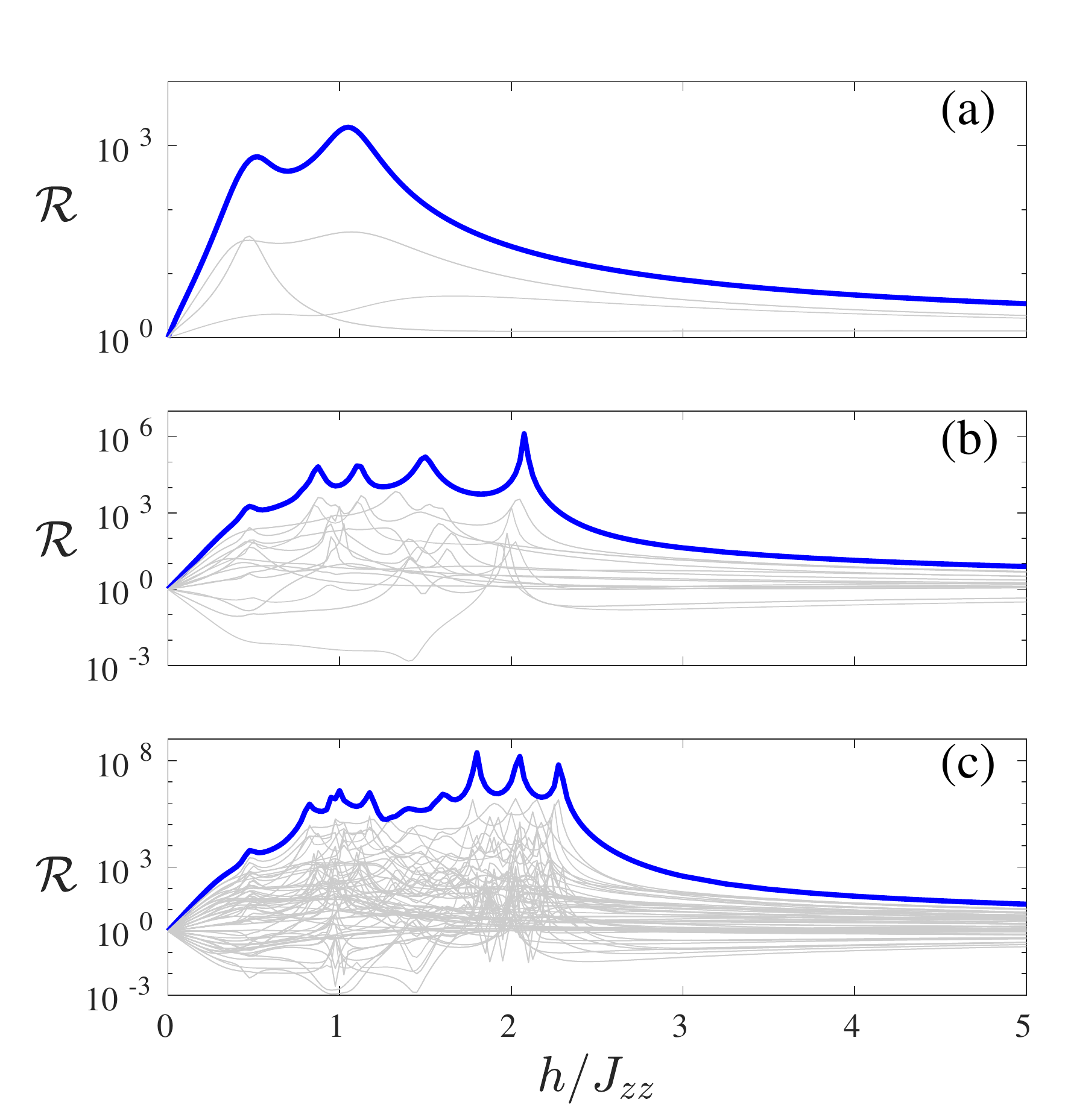}
\caption{Rectification $\mathcal{R}$ is plotted as a function of the ratio of local field strength and interaction $h/J_{zz}$ for system sizes $L=4$ (a),  $L=6$ (b) and $L=8$ (c). The current $\mathcal{J}_f$ with magnetic field configuration as $(+,\dots, +, -, \dots, -)$ and $\mathcal{J}_r$ for $(-,\dots, -, +, \dots, +)$ are highlighted as thick blue lines. The other configurations are in thin grey lines. Common parameters are $J_{zz}=4$, $\gamma=1$, $\mu=1$.}
\label{fig:rectification}
\end{figure}

We thus investigate the rectification in Fig.\ref{fig:rectification}. Here, the rectification is quantified using $\mathcal{R}=\mathcal{J}_f/\mathcal{J}_r$ \cite{TerraneoCasati2002, LiCasati2004, LiLi2012, balachandran_perfect_2018} where $\mathcal{J}_f$ and $\mathcal{J}_r$ are referred to as forward and reverse currents and are computed, respectively, for a configuration of the magnetic field and its reflection symmetric one, e.g. $(+,-,+,-)$ and $(-,+,-,+)$. We note that this is equivalent to fixing a configuration and switching the driving bias (i.e. $\mu=1$, forward direction to $\mu=-1$, reverse direction). When $\mathcal{R}=1$, there is no rectification as the forward and reverse currents are equal, e.g. for symmetric magnetic fields configurations. Perfect diodes are signalled by $\mathcal{R}=\infty$ or $0$ (the latter is obtained when the forward current tends to $0$ but the reverse current is finite).
In Fig.\ref{fig:rectification} there are less lines compared to Figs.\ref{fig:currentvsdelta}-\ref{fig:currentvsh}, and this is due to the fact that each line corresponds to a pair of magnetic field configurations: one is a configuration, and the other is the reflection symmetric one. Importantly, each pair is considered only once, e.g. we plot the rectification considering the $(+,-,+,-)$ configuration to give the forward current $\mathcal{J}_f$ and $(-,+,-,+)$ to give the reverse current $\mathcal{J}_r$, and we do not plot the opposite combination because it does not give extra information, resulting in a $1/\mathcal{R}$ rectification. This is particularly relevant because in Fig.\ref{fig:rectification} we use a log-lin plot, and the reverse combination of magnetic field configurations would simply result in a curve symmetric around $\mathcal{R}=1$.
The blue thick line in Fig.\ref{fig:rectification} corresponds to the combination $(+,\dots, +,-, \dots, -)$, for $\mathcal{J}_f$, and $(-,\dots, -, +, \dots, +)$, for $\mathcal{J}_r$, and it gives clearly the strongest rectification. We remind the reader that a small value of $\mathcal{R}$ correspond to a large rectification in the opposite direction, yet clearly the blue thick line corresponds to the largest possible current rectifications. In Fig.\ref{fig:rectification} we also note that for larger systems one can obtain even larger rectifications, for example showing a rectification of $\mathcal{R}\approx 10^8$ for the $L=8$ chain. We will also return to this point in a later part of the paper.

\begin{figure}[ht]
    \includegraphics[width=\columnwidth]{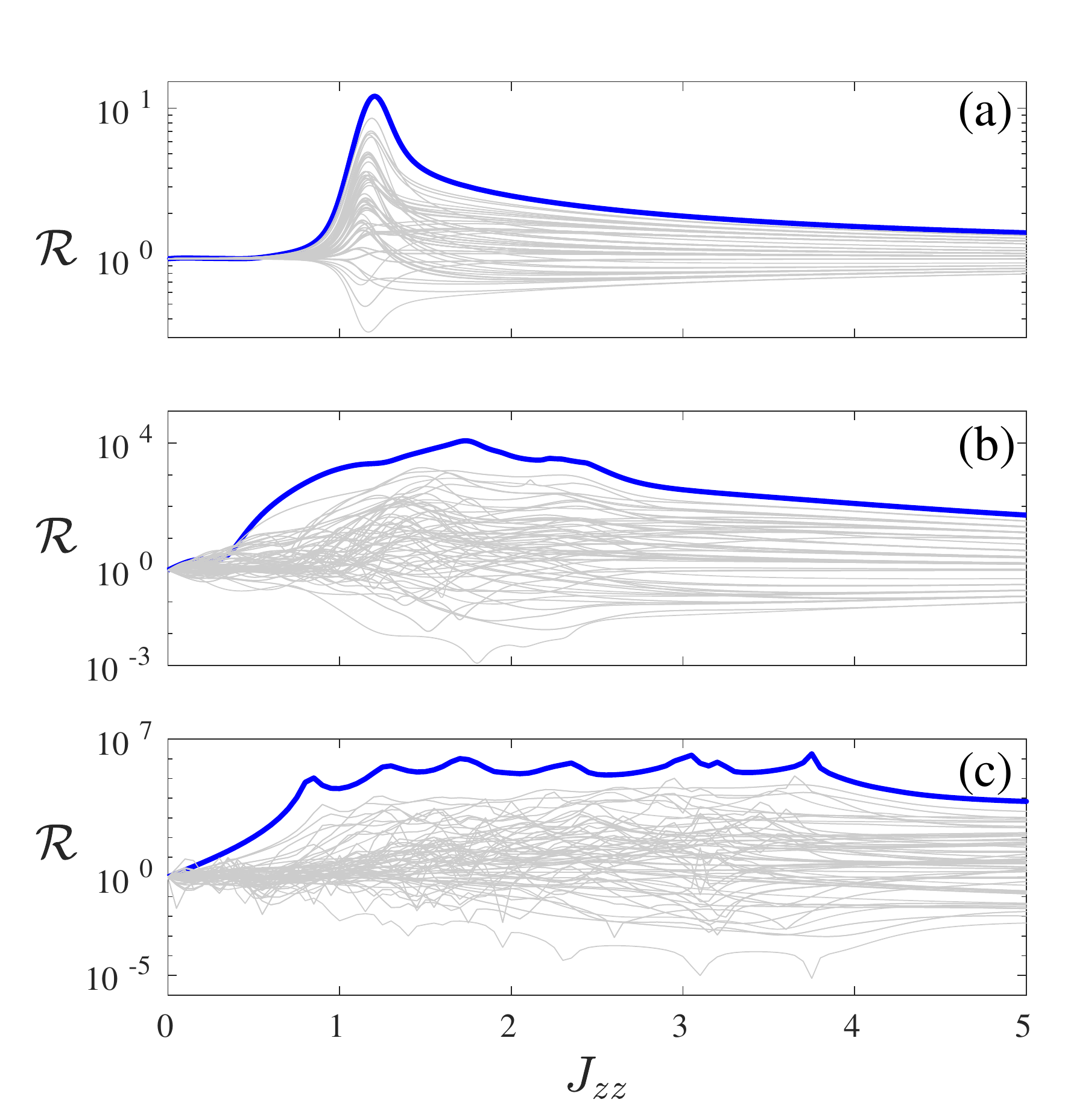}
\caption {Rectification $\mathcal{R}$ is plotted as a function of interaction $J_{zz}$ for $h=0.1$ (a), $h=1$ (b) and $h=3$ (c) for a system size of $L=8$. The rectification as ratio of current $\mathcal{J}_f$ with magnetic field configuration as $(+,\dots, +, -, \dots, -)$ and $\mathcal{J}_r$ for $(-,\dots, -, +, \dots, +)$ are highlighted as thick blue lines. The other configurations are in thin grey lines. Common parameters are $\gamma=1$, $\mu=1$.}
\label{fig:rvsjzz}
\end{figure}

%\khc{
In Fig.\ref{fig:rvsjzz}, rectification $\mathcal{R}$ is plotted as a function of interaction $J_{zz}$. Similar to Fig.\ref{fig:rectification}, each line corresponds to a pair of magnetic field configurations which are the reflection symmetric of each other. Highlighted in blue thick line is the $(+,\dots +, -, \dots -)$, $(-,\dots, -, +, \dots, +)$ configuration pair which yields the strongest rectification. Here, we highlight the role of interaction $J_{zz}$ in causing large rectification. In panel (a) of Fig.\ref{fig:rvsjzz} where $h=0.1$, we observe the sharp transition to a steep increase in rectification occuring near $J_{zz}=1$, where the quantum phase transition occurs. This transition occurs at smaller values of $J_{zz}$ for increasing $h$ as we observe for $h=1$ in panel (b) and $h=3$ in panel (c). With increasing $h$, the system behaviour deviates further from that of the XXZ spin chain system, and it is thus natural that the values of $J_{zz}$, for which an enhancement of rectification occur, deviates further from $J_{zz}=1$. Fig.\ref{fig:rvsjzz} thus highlights the importace of the interplay of kinetic, interactive and dissipative terms in the master equation (\ref{lindblad}) of the set-up.   
%the $0-$current-carrying state $|UD\rangle$ gets increasingly favoured by the dissipator in the reverse bias and the state $|DU\rangle$ gets increasingly favoured in the forward bias. Thus, the system is able to gain rectification power with smaller interaction strengths when local field strength increases.}

\begin{figure}[ht]
    \includegraphics[width=\columnwidth]{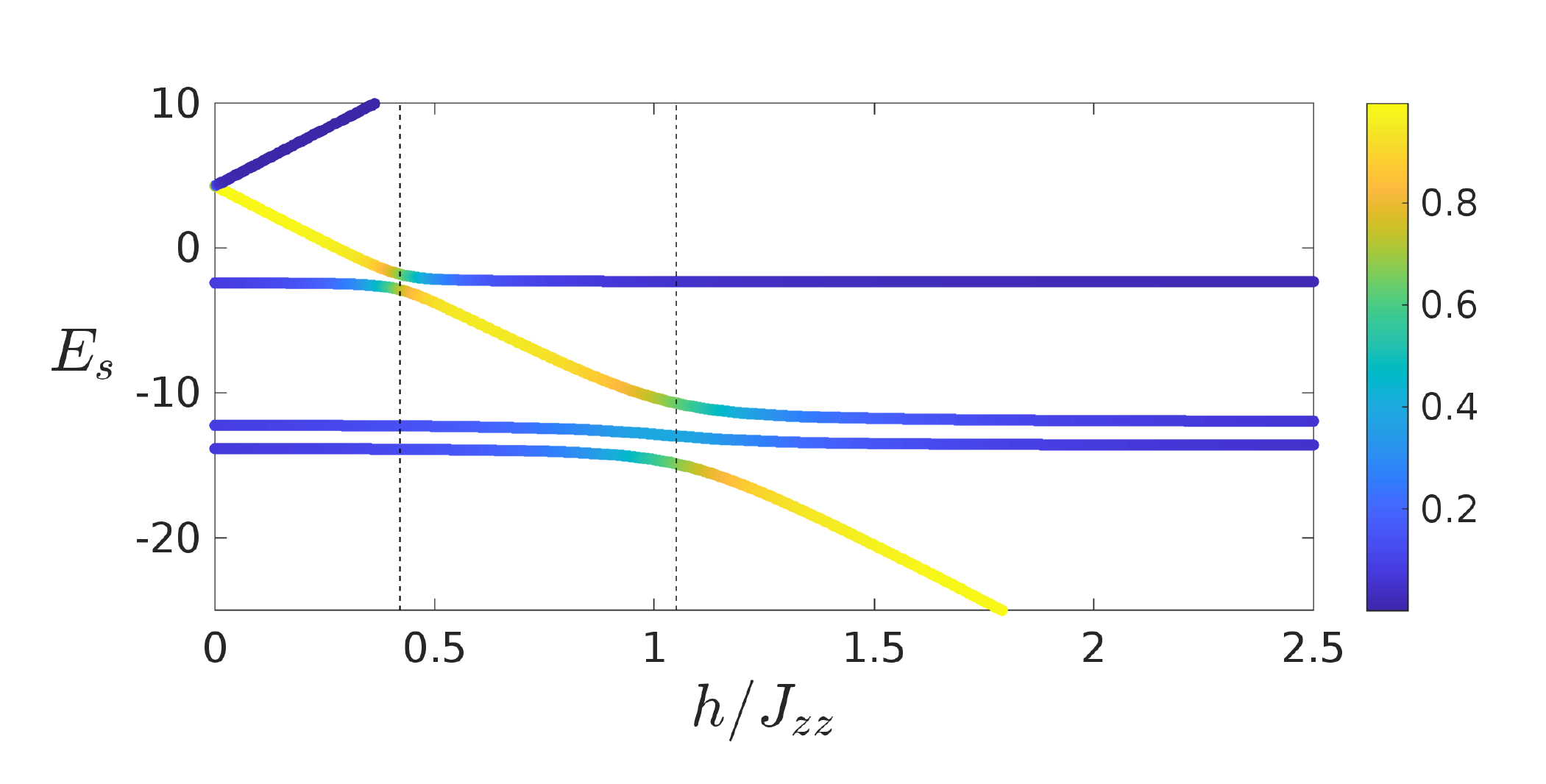}
\caption {Eigenenergies $E_s$ in the zero magnetisation sector of a chain of length $L=4$ with magnetic field configuration $(+,+,-,-)$ plotted as a function of the ratio of local field strength and interaction $h/J_{zz}$. The vertical lines in each panel correspond to peaks in the current as from Fig.\ref{fig:currentvsh}(a). The color used for the eigenenergies corresponds to the overlap between the eigenvector and the state $|\!\!\downarrow\downarrow\uparrow\uparrow\rangle$. Parameters: $\gamma=1$, $\mu=1$, $J_{zz}=4$.}
\label{fig:Fig5}
\end{figure}

In Figs.\ref{fig:currentvsdelta}-\ref{fig:rvsjzz} we have observed resonances, which correspond to peaks of currents and the largest rectifications. We now aim to gain an insight into this.
The mechanism for the emergence of such resonances, and of the strong rectifications, can be understood by studying the configuration $(+,\dots +, -, \dots -)$, where the field is in the positive direction in first half of the chain and negative in the other half of the chain (the configuration corresponding to the largest rectification).
To give a clear idea of the mechanism we focus on the case of $L=4$. In Fig.\ref{fig:Fig5} we show the energy spectrum as a function of $h$ for $J_{zz}=4$. For each magnitude $h$, the value of the energy is indicated by a point in the plot. Different colors of each point corresponds to the values of the overlap of the corresponding eigenvector $|\psi_s\rangle$ with the state $|\!\downarrow\downarrow\uparrow\uparrow\rangle$, i.e. $|\langle\psi_s|\!\downarrow\downarrow\uparrow\uparrow\rangle|^2$. The vertical dashed lines  shows the position of the peaks of current for system size $L=4$, as taken from Fig.\ref{fig:currentvsh}. It is clear from the figure that the avoided crossings in the spectrum matches with the maxima of the current. At these points, given the proximity in energy between different energy eigenstates, it is easier for the steady state to be in a mixture of different eigenstates, thus resulting in the possibility of larger currents (note that each energy eigenstate carries $0$ current). The presence of avoided crossings for the peaks in Figs.\ref{fig:currentvsdelta}-\ref{fig:rectification} has been checked for all system sizes and parameters tested.

\begin{figure}[ht]
    \includegraphics[width=\columnwidth]{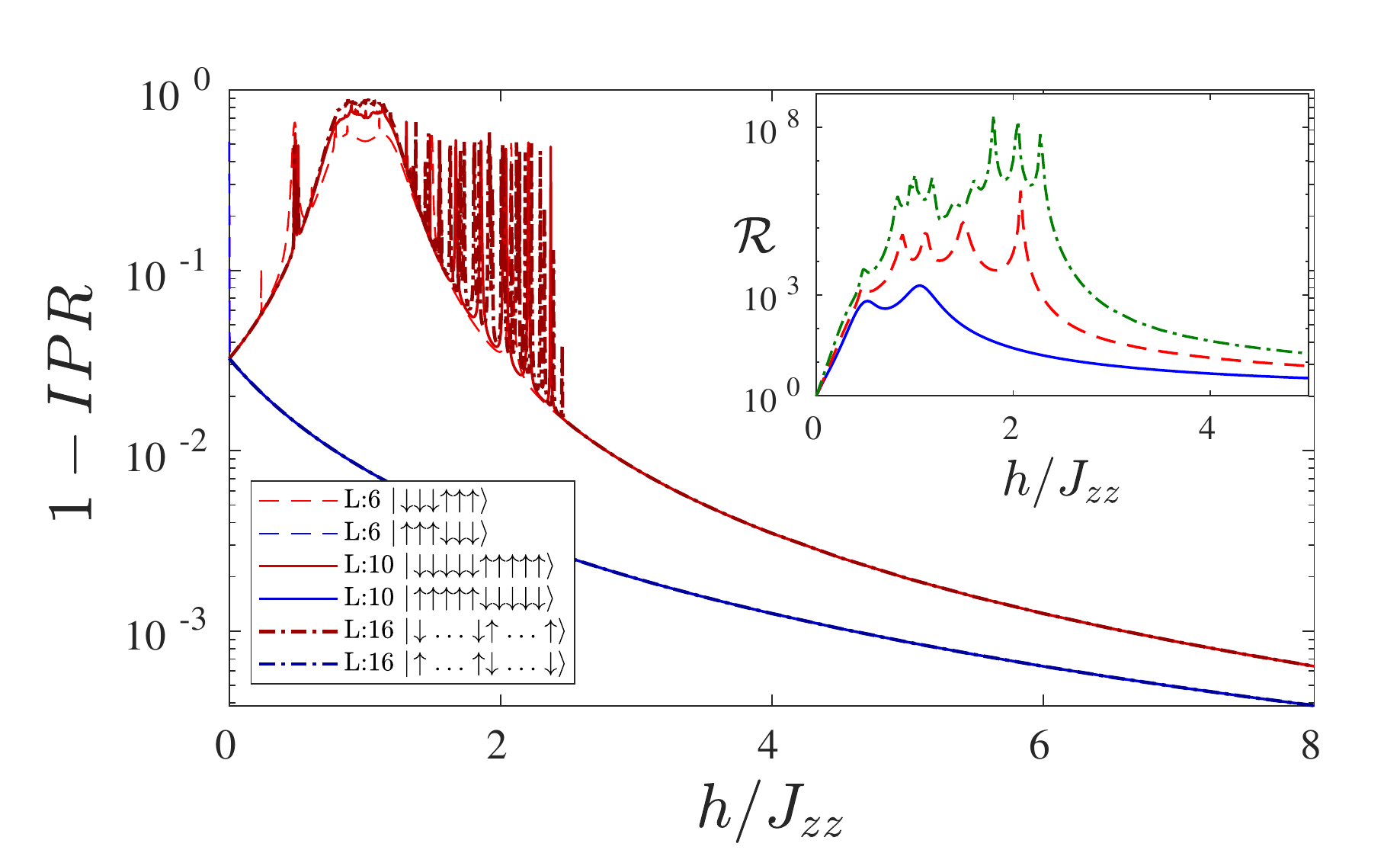}
\caption{The inverse participation ratio $1-IPR$ as a function of the ratio of local field strength and interaction $h/J_{zz}$ for different system sizes $L=16$ (dashed line), $L=10$ (continuous line) and $L=6$ (dot-dashed line). Here the magnetic field is in the configuration $(+,\dots,+,-,\dots,-)$. Both state configurations $|DU\rangle=|\!\downarrow\dots\downarrow\uparrow\dots\uparrow\rangle$ (red) and $|UD\rangle=|\!\uparrow\dots\uparrow\downarrow\dots\downarrow\rangle$ (blue) are shown. The inset shows the rectification as a function of $h/J_{zz}$ for different system sizes: blue continuous line for $L=4$, red dashed line for $L=6$ and green dot-dashed line for $L=8$. Other parameters are $J_{zz}=4$, $\gamma=1$, $\mu=1$.}
\label{fig:iprvsh}
\end{figure}

We have thus shown that an XXZ chain with large enough interaction $J_{zz}$ and a magnetic field in the configuration $(+,\dots +,-,\dots -)$ results in a highly performing spin-current diode. It is however important to investigate the performance at larger sizes of this diode. 
At this point we should stress that computing the steady state in regimes of very low currents is extremely demanding because the equations are ill-conditioned. We thus resort to a different, yet equally insightful approach.   
To understand the robustness of the effect for larger system sizes, we study the inverse participation ratio ($IPR$) of the Hamiltonian of the system with the local magnetic field configuration $(+,\dots, +,-,\dots, -)$ in Fig.\ref{fig:iprvsh}.
The $IPR$ for a given state $|\psi\rangle$ over the energy eigenstates $|n\rangle$ is given by $IPR=\sum_n |\langle n|\psi\rangle|^4 $. A value of $1-IPR \approx 1$ means that the state $|\psi\rangle$ is well distributed over all the eigenstates $|n\rangle$, while $1-IPR \approx 0$ means that $|\psi\rangle$ almost exactly corresponds to a single energy eigenstate. The study of this quantity can be done simply by diagonalizing the Hamiltonian, which we do for system sizes up to $L=16$.

In Fig.\ref{fig:iprvsh}, we plot $1-IPR$ as a function of the ratio of local field strength and interaction $h/J_{zz}$ for the state $|DU\rangle=|\!\downarrow\dots\downarrow\uparrow\dots\uparrow\rangle$ (red plots) and for the state $|UD\rangle=|\!\uparrow\dots\uparrow\downarrow\dots\downarrow\rangle$ (blue plots). For the state $|UD\rangle$, $1-IPR$ quickly becomes small and it continues to decrease as $h$ increases. This means that the state which is favored by the dissipator in reverse bias, $|UD\rangle$, is almost entirely an eigenstate of the Hamiltonian. Hence the steady state would be well approximated by this $0-$current-carrying state. We note that the blue solid, dashed and dot-dashed curves relative to this state, each for a different system size, are almost completely identical.

For the state $|DU\rangle$ the physics is very different. For $h=0$ it approximates well the highest energetic state, together with $|UD\rangle$. But while $|UD\rangle$ approximates better and better the highest energetic state as $h$ increases, for large enough magnetic field $h$, the state $|DU\rangle$ well approximates the ground state. Hence this state is bound to go through numerous avoided crossings, at the occurrence of which transport is favored and rectification can be very large.  
In particular, we observe that for the state $|DU\rangle$, $1-IPR$ is close to $1$ for $h \approx J_{zz} $, and $1-IPR$ increases with the system sizes. 
This is because of the presence of a energy band of state which are crossed for $h \approx J_{zz} $. Beyond this energy band there can be other avoided crossings which can result in even larger rectification. For instance, in the inset of Fig.\ref{fig:iprvsh} we illustrate the rectification for different system sizes, showing a significant increase in the rectification power as the system size increases, even up to $\mathcal{R}=10^8$ ($L=4$ blue continuous line, $L=6$ red dashed line and $L=8$ green dot-dashed line). The exact position of the last avoided crossings depends on the parameters of the system. For large enough $J_{zz}/J$, they occur for $h\approx 2J_{zz}$. This can be computed analytically, in fact setting $J=0$ one realizes that the energy of state $|DU\rangle$ is, for large enough system sizes $L$, given by $(L-3)J_{zz}-L h$ while the energy of the first excited state (at large enough $h$), is $(L-11)J_{zz}-(L-4) h$. These two energies coincide for $h=2J_{zz}$. For finite values of $J$, and smaller ratios $J_{zz}/J$ the last resonance is moved to larger values of $h$, as in the cases analyzed in this work.

%%%%%%%%%%%%%%%%%%%%%%%%%%%%%%%%%%%%%%%%%%
\section{Conclusion}\label{sec:conclusions}
We have studied the effect of local magnetic fields on the spin transport of a strongly interacting XXZ chain.
We have shown that a configuration with the field pointing in one direction for half the chain and in the opposite direction for the other half is particularly effective to produce strong rectification. This is due to the fact that in one direction the magnetic field cooperates with interactions in producing two large ferromagnetic domains, while in the other direction the magnetic field opposes such formation and favors transport. Rectification is particularly enhanced at the occurrence of avoided crossings in the energy spectra of the XXZ chain with this configuration of magnetic field.
We show giant rectification, for instance, for $L=8$ we show rectifications of the orders of $10^8$. We also show that the rectification is not only robust upon increasing the chain length, but strengthened. Moreover, the presence of resonant peaks of rectification can turn this setup into a switch or sensor, activated by small changes in the magnetic field.

In future works we could consider the stability of this effect against other forms of dissipation. For instance, dephasing has been shown to suppress negative differential conductance \cite{MendozaArenasClark2013}, or perturbations such as long range interactions have a detrimental effect on negative differential conductance \cite{DroennerCarmele2017}. Another possible direction would be to consider the performance of this diode for heat current rectification \cite{TerraneoCasati2002, LiCasati2004, LiLi2012}.

\begin{acknowledgments} 
D.P. acknowledges support from the Ministry of Education of Singapore AcRF MOE Tier-II (Project No. MOE2016-T2-1-065). This work was partially performed at and supported by the MPI-PKS Advanced Study Group ``Open quantum systems far from equilibrium''. The computational work for this article was performed on resources of the National Supercomputing Centre, Singapore \cite{nscc}.
\end{acknowledgments}

%%%%%%%%%%%%%%%%%%%%%%%%%%%%%%%%%%%%%%%%%%
%\conflictsofinterest{The authors declare no conflict of interest.} 

%%%%%%%%%%%%%%%%%%%%%%%%%%%%%%%%%%%%%%%%%%
%% optional
%\abbreviations{The following abbreviations are used in this manuscript:\\

%\noindent 
%\begin{tabular}{@{}ll}
%NDC & Negative Differential Conductivity\\
%GKSL & Gorini-Kossakowski-Sudarshan-Lindblad\\
%NESS & Non-equilibrium steady state\\
%\end{tabular}}


%=====================================
% References, variant A: internal bibliography
%=====================================
%\reftitle{References}
\bibliography{disorderupdate}

% The following MDPI journals use author-date citation: Arts, Econometrics, Economies, Genealogy, Humanities, IJFS, JRFM, Laws, Religions, Risks, Social Sciences. For those journals, please follow the formatting guidelines on http://www.mdpi.com/authors/references
% To cite two works by the same author: \citeauthor{ref-journal-1a} (\citeyear{ref-journal-1a}, \citeyear{ref-journal-1b}). This produces: Whittaker (1967, 1975)
% To cite two works by the same author with specific pages: \citeauthor{ref-journal-3a} (\citeyear{ref-journal-3a}, p. 328; \citeyear{ref-journal-3b}, p.475). This produces: Wong (1999, p. 328; 2000, p. 475)

%=====================================
% References, variant B: external bibliography
%=====================================
%\externalbibliography{yes}
%\bibliography{your_external_BibTeX_file}

%%%%%%%%%%%%%%%%%%%%%%%%%%%%%%%%%%%%%%%%%%
%% optional
% \sampleavailability{Samples of the compounds ...... are available from the authors.}

%% for journal Sci
%\reviewreports{\\
%Reviewer 1 comments and authors’ response\\
%Reviewer 2 comments and authors’ response\\
%Reviewer 3 comments and authors’ response
%}

%%%%%%%%%%%%%%%%%%%%%%%%%%%%%%%%%%%%%%%%%%

\end{document}